\documentclass[12pt]{iopart}

\usepackage{graphicx}% Include figure files

\usepackage{amssymb}
\usepackage{bbm}

\usepackage[OT4,T1]{fontenc}
\usepackage[utf8]{inputenc}

\newcommand{\mket}[1]{| #1 \rangle}

\newcommand{\mbraket}[2]{\langle #1 | #2 \rangle}

\newcommand{\imag}{\mathbbm{i}}

\begin{document}

\title[Quantum circuits based on qutrits as a tool for solving systems of linear equations]{Quantum circuits based on qutrits as a tool for solving systems of linear equations}

\author{Marek Sawerwain}

\address{
Institute of Control and Computation Engineering \\ University of Zielona G\'ora, ul. Licealna 9, Zielona G\'ora 65-417, Poland}
\ead{M.Sawerwain@issi.uz.zgora.pl}

\author{Wiesław Leoński}

\address{
Quantum Optics and Engineering Division, Institute of Physics \\ University of Zielona G\'ora, ul. Licealna 9, Zielona G\'ora 65-417, Poland}
\ead{W.Leonski@proton.if.uz.zgora.pl}

\begin{abstract}
Recently, it has been presented some algorithms and physical models which give prospects for construction of quantum computers capable to solve systems of linear equations. The common feature which is shared in these works is the use of qubits which allow to solve systems with $2^n$ variables. In this work we propose a~quantum circuit based on qutrits architecture which directly allows for solving systems of equations with three variables. Proposed circuit can be easily generalized to those  with $3^n$ variables. We also present some numerical experiments to verify the correctness of proposed solution.
\end{abstract} 

\vspace{2pc}

\maketitle

\section{Introduction}

Solving system of linear equations (SLE) is common computational task in modern science, technology and engineering. Although classical algorithm for solving of SLE is of polynomial complexity, the increasing size of SLE causes the bigger processing time, which naturally provides a strong motivation for looking for better and faster methods of solving such problems.

In \cite{Harrow2009} authors proposed a quantum algorithm for solving SLE (qaSLE), which offers exponential speed-up to any known today classical algorithms for solving SLE. What is important, qaSLE uses only linear amount of quantum resources to solve SLE. Recently, some proposals of experimental realization of qaSLE were shown and discussed \cite{Barz2013, Cai2013, Cao2012}.
In those papers the system of linear equations is represented by qubits. This fact implies some difficulty with the construction of circuits for solving of SLE of odd dimensions. In consequence, it is not possible to construct a circuit for matrix $A$ of $3 \times 3$ dimension directly. In this paper we propose approach where the qutrits are used for direct construction of suitable quantum circuit which can be used to solve SLE with three variables.

This paper is organized as follows. Section~\ref{lbl:sec:qunatum:alg:for:SLE:solving} briefly presents the algorithm based on the ideas discussed in the paper~\cite{Harrow2009} for solving SLE with use of qutrits instead of qubits is briefly presented. In section~\ref{lbl:sec:qutrits:circuits:for:SLE:solving} the definition of gates necessary for construction of our circuit are given, and the form of proposed quantum circuit for solving SLE is presented. Section~\ref{lbl:sec:num:experiments} presents numerical simulations and their results, corresponding to the process of searching of solution. The conclusion and remarks will be presented in section~\ref{lbl:sec:conclusions}.

\section{Quantum algorithm for solving system of linear equations} \label{lbl:sec:qunatum:alg:for:SLE:solving}

In \cite{Harrow2009} the problem of solving of SLE is represented by a main matrix $A$, vector of unknown variables $X$ and vector of free values $B$. This problem can be expressed by the relation $A \mket{X} = \mket{B}$, where $\mket{X}$ and $\mket{B}$ are normalized quantum states which represents the vector of unknown variables and free values, respectively.  It is assumed that $A$ is $N \times N$ Hermitian matrix, where $N$ denotes the total number of variables appearing in the system, whereas $\{ \mket{u_j} \}$ forms a base of $A$. Additionally, eigenvalues $\{ \lambda_j \}$ fulfill the condition $0 < \lambda_j < 1$.
The base of the matrix $A$ is also used to express the state $\mket{B}$, and unknown vector $\mket{X}$ representing the solution. Hence, we have:
\begin{equation}
\mket{B} = \sum_{j=0}^{N-1}  \beta_j \mket{u_j}, \;\;\;  \mket{X} = \frac{ A^{-1} \mket{B}}{|| A^{-1} \mket{B} ||},
\label{lbl:eqn:free:values:and:solutions}
\end{equation}
where coefficients $\beta_j = \mbraket{u_j}{B}$, whereas $A^{-1}$ represents the inverse of matrix $A$.

Quantum algorithm for solving SLE described in \cite{Harrow2009} for system $A \mket{X} = \mket{B}$ can be summarized in four steps: phase estimation,  ''non-unitary'' operation (implemented by the controlled rotation) on ancilla qubit, reverse phase estimation and finally, measurement of an ancillary qubit/qutrit.

In this article we propose a quantum circuit for solution of SLE with three variables that is based on qutrits. It can be defined by
\begin{equation}
\mket{\psi} = \sum_{j=0}^{2} \alpha_j \mket{j}, \;\;\; \alpha_i \in \mathbb{C},
\end{equation}
where the states $\mket{0}, \mket{1}, \mket{2}$ define a standard computational base for our qutrit system.

Irrespectively to the fact whether the qubit or qutrit (or even qudit in general case) system is considered, the solution of SLE can be written as:
\begin{equation}
\mket{X} = \sum_{j=1}^{N} \beta_j \frac{1}{\lambda_j} \mket{u_j}
\label{lbl:eq:solution}
\end{equation}
To obtain the solution we have to start with the qutrit input state ($N=3$) which represents the free values defined in eq.~(\ref{lbl:eqn:free:values:and:solutions}) and denoted as $\mket{B}$.
In consequence, the quantum operations of phase estimation gives the following state
\begin{equation}
\mket{\psi_{1}} = \sum_{j=1}^{N} \beta_{j} \mket{u_j} \mket{\lambda_j},
\end{equation}

In the second step, additional rotation necessary for simulation of an non-unitary operation transforms the $\mket{\psi_{1}}$ into
\begin{equation}
\mket{\psi_{2}} = \sum_{j=1}^{N} \beta_{j} \mket{u_j} \mket{\lambda_j} \mket{s_j},
\end{equation}
where 
\begin{equation}
\mket{s_j} = \left( \sqrt[3]{2-\frac{C^3}{\lambda^{3}_{j}}} \mket{0} + \sqrt[2]{1-\frac{C^2}{\lambda^{2}_{j}}} \mket{1} + \frac{C}{\lambda_{j}} \mket{2} \right),
\label{lbl:eq:sj:state:v1}
\end{equation}
and $C \leq \min|\lambda_j|$. The solution of SLE can be found in the component $\frac{C}{\lambda_{j}} \mket{2}$. One should keep in mind that the state (\ref{lbl:eq:sj:state:v1}) is not normalized. Therefore, normalization should be performed and we get:
\begin{equation}
\mket{s_j} = \left( \sqrt{1-\frac{C^2}{\lambda^{2}_{j}}} \mket{0} + \frac{C}{\lambda_{j}} \mket{2} \right) .
\end{equation}
Additionally, the elimination of state $\mket{1}$ increases the probability of measure of the state $\mket{2}$ which is performed in the third step.
Therefore, before the final measurement, the reverse phase estimation is used to find the state $\mket{\lambda_j}$ and finally, we perform measurement the thrid ancillary qutrit.
If in the process of computation the state $\mket{2}$ is measured, the final state becomes
\begin{equation}
\mket{\psi_{3}} = \sum_{j=1}^{N} C \frac{\beta_{j}}{\lambda_{j}} \mket{u_j}.
\end{equation}
This state represents normalised solution given by eqn.~(\ref{lbl:eq:solution}).

\section{Quantum qutrit circuit for SLE solving}
\label{lbl:sec:qutrits:circuits:for:SLE:solving}

\begin{figure}
\begin{center} 
\includegraphics[height=0.15\columnwidth]{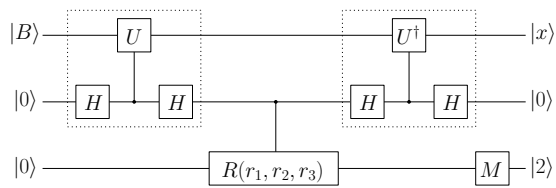} 
\end{center}
\caption{Quantum circuit based on qutrits applied for solving SLE with three unknown variables. Blocks labelled by  $H$ denote Hadamard gates, $U$ gates for phase estimation, and $R$ rotation gates defined in (\ref{lbl:eq:R:form:one}).}
\label{lbl:fig:quantum:circuit:for:solution}
\end{figure}

What is necessary to fully define proposed circuit, apart of the phase estimation and rotation gates definitions, is to define Hadamard gate (HG) within general qudit formalism:
\begin{equation}
H \mket{j} = \sum_{k=0}^{2}\omega^{j \cdot k}\mket{k}, \quad \omega^3_k = \cos \left( \frac{2k\pi}{3} \right) + \imag \sin\left(\frac{2k\pi}{3}\right) = e^{\frac{2\pi \imag k}{3}},
\end{equation}
where $k = \{0, 1, 2\}$, $\omega$ represents the $k$-th primitive root of unity and $\imag$ symbol denotes imaginary unit. The rotation gate for qutrits is defined by $R(\theta) = e^{\imag \theta_i / 3}$ where operators $\theta_i$ have a form:
\begin{equation}
\begin{array}{lll}
\theta_2 = \left(
\begin{array}{ccc}
 0 & -\imag & 0 \\
 \imag & 0 & 0 \\
 0 & 0 & 0
\end{array}
\right), &
\theta_5 = \left(
\begin{array}{ccc}
 0 & 0 & -\imag \\
 0 & 0 & 0 \\
 \imag & 0 & 0
\end{array}
\right), &
\theta_7 = \left(
\begin{array}{ccc}
 0 & 0 & 0 \\
 0 & 0 & -\imag \\
 0 & \imag & 0
\end{array}
\right) 
\end{array}
\label{lbl:gell:mann:matrices}
\end{equation}
and came form the set of so-called Gell-Mann's operators. They allow for much easier formulation of qutrit rotation gates applied in simulation of the step in which non-unitary operation is performed. Thus, the definition of a qutrit rotation gate $R$ used in our quantum circuit and corresponding to the solving SLE with three variables can be written as follows:
\begin{equation}
\begin{array}{lll}
r_1 = -2.0 \arccos( l_1 ), & R_1  = e^{\imag r_1 \theta_2 / 3}  & \\
r_2 = -2.0 \arccos( l_2 ), & R_2  = e^{\imag r_2 \theta_5 / 3}, &  R = R_1 \cdot R_2 \cdot R_3 \\
r_3 = -2.0 \arccos( l_3 ), & R_3  = e^{\imag r_3 \theta_7 / 3}  & \\
\end{array}
\label{lbl:eq:R:form:one}
\end{equation}
The use of $\arccos$ function allows for change of the domain of $l_1$, $l_2$, $l_3$ parameters from the  range $\langle -\pi, \pi \rangle$ to  $\langle-1, 1 \rangle$.
The use of gate $R$ (see the circuit in Fig.~(\ref{lbl:fig:quantum:circuit:for:solution})) enable us find the solution of a given SLE. 

What is interesting, the gate $R$ can be built also only with use of the operator $\theta_5$:
\begin{equation}
R_1  = e^{\imag r_1 \theta_5 / 3}, R_2  = e^{\imag r_2 \theta_5 / 3}, 
R_3  = e^{\imag r_3 \theta_5 / 3}, R = R_1 \cdot R_2 \cdot R_3 .
\label{lbl:eq:R:form:two}
\end{equation}
since for this case state $\mket{s_j}$ can be approximated by:
\begin{equation}
\mket{s_j} = \cos(r_j) \mket{0} + \sin(r_j) \mket{2},
\end{equation}
where $\sin{r_j} \approx \frac{C}{\lambda_j}$ and $r_j = \arcsin(\frac{C}{\lambda_j})$.

The phase estimation is represented by operator 
$U = e^{2 \pi \imag 3^{n} A }$, where $n$ denotes the position of the first digit (in the case of qutrit the ternary digit) which allows for differentiation of eigenvalues, \textit{e.g.} for ternary numbers $0.221$, $0.220$, $0.222$ it is third position.

\section{Numerical experiments}
\label{lbl:sec:num:experiments}
To validate our ideas we shall perform some numerical experiment. Thus, we assume that the main matrix $A$ of a system of linear equations is given by:
\begin{equation}
A = \left(
{
\begin{array}{lll}
0.44033 + 0.00000\imag  &  0.05719 - 0.02612\imag   & 0.02565 + 0.05151\imag \\
0.05719 + 0.02612\imag  &  0.40686 + 0.00000\imag   & 0.05915 + 0.00073\imag \\
0.02565 - 0.05151\imag  &  0.05915 - 0.00073\imag   & 0.48614 + 0.00000\imag
\end{array}
} \right)
\label{lbl:eqn:matrix:A}
\end{equation}
whereas free values and eigenvalues are:
\begin{equation}
B = \left(
{
\begin{array}{l}
0.56751 \\
0.79592 \\
0.21084
\end{array}
} \right), \lambda_0 = 0.33(3), \lambda_1 = 0.44(4), \lambda_2 = 0.55(5)
\label{lbl:eqn:free:values:A}
\end{equation}
For our purposes given eigenvalues were initially denoted in ternary system \textit{i.e.} $\lambda_0 = (0.10)_{3} = (0.333(3))_{10}$, $\lambda_1 = (0.11)_{3} = (0.444(4))_{10}$, $\lambda_2 = (0.12)_{3} = (0.555(5))_{10}$. It is seen that here the second digit discriminates given eigenvalues. Therefore, the parameter $n$ is assumed to be equal to two, when the phase estimation is performed.

\begin{figure}
\begin{center} 
\begin{displaymath}
\begin{array}{cccc}
(a) & \includegraphics[width=0.35\columnwidth]{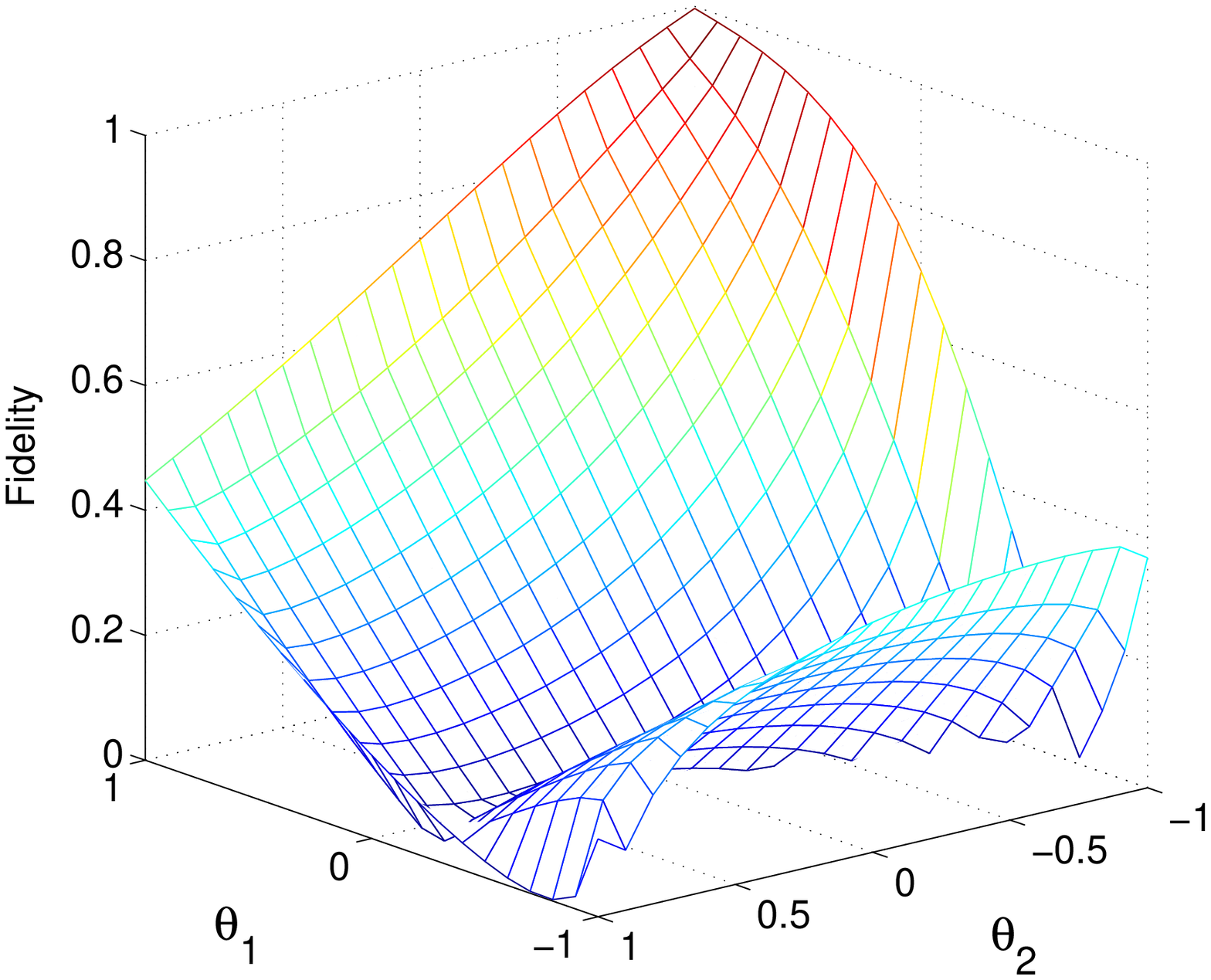} & (b) & \includegraphics[width=0.35\columnwidth]{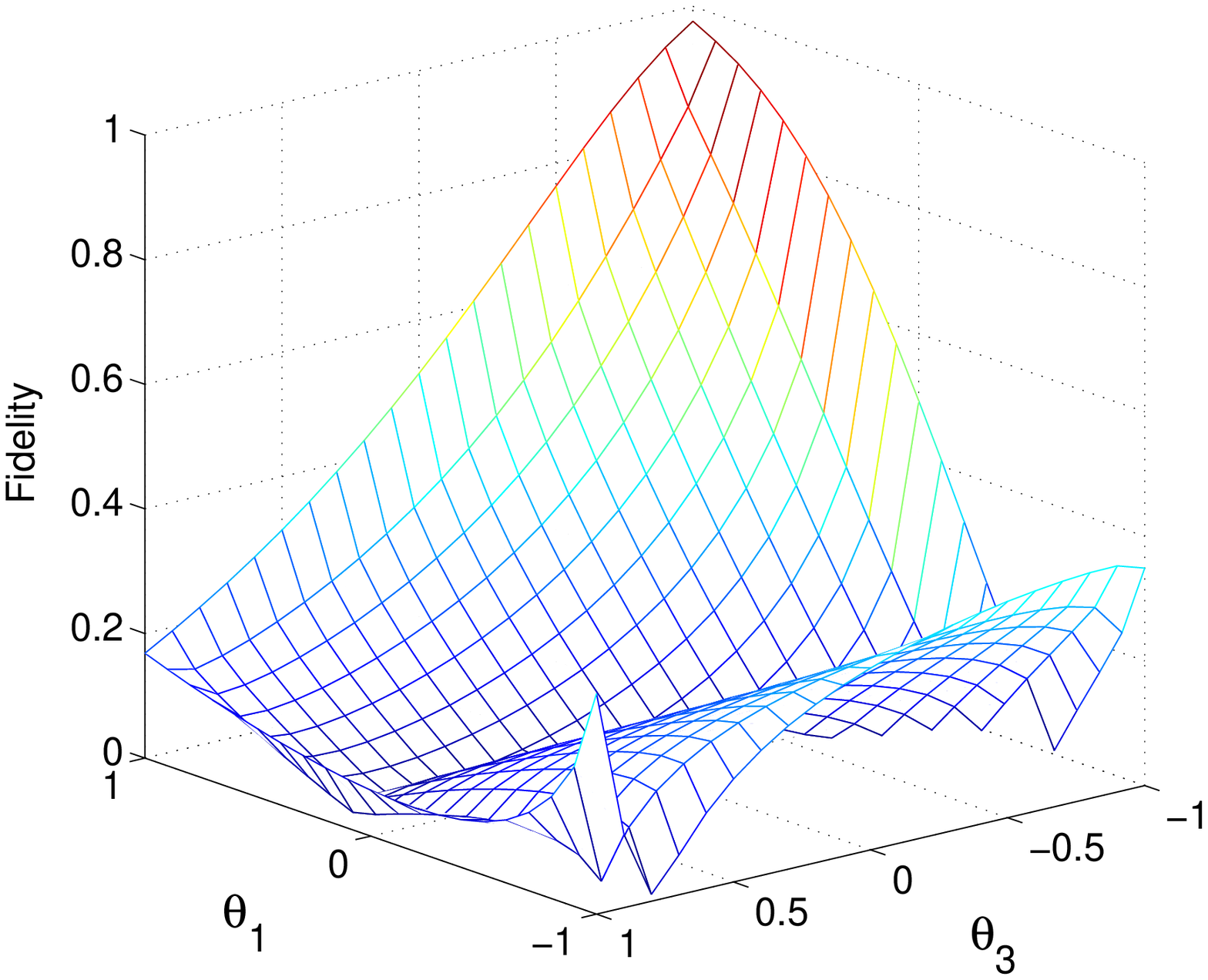}  
\end{array}
\end{displaymath}
\end{center}
\caption{The fidelity between a proper solution and the final state of proposed quantum circuit. Case (a):  $r_3=0$ and $r_1$, $r_2$ are changing from $-1$ to $1$,  case (b):  $r_2=-.25$  and $r_1$, $r_3$ are changing within the same range as $r_1$, $r_2$ in case (a). }
\label{lbl:fig:searing:for:solution}
\end{figure}

In practice, the eigenvalues are not known. To solve particular SLE we need to find all parameters determining the rotation gates and the parameter $n$ which give the solution of SLE.  If we assume that $n=2$, by changes of the values of $r_1$, $r_2$, $r_3$ one can find solution with sufficiently high fidelity, what is illustrated in Fig.\ref{lbl:fig:searing:for:solution}. We see that it is possible to achieve value of the fidelity equal to $0.99$ when  $r_1 = -1$, $r_2=1$, $r_3 = 0.25$. For the case considered here, finding values of the parameters $r_1$, $r_2$ plays a crucial role in the process of searching  solution, whereas the parameter $r_3$ can be applied to improve the accuracy of final solution. Finally, for the example presented here, the obtained state that represents the normalized solution of SLE given by eqns.~(\ref{lbl:eqn:matrix:A}), (\ref{lbl:eqn:free:values:A}) and is equal to:
\begin{equation}
X = \left(
{
\begin{array}{l}
 0.508890 + 0.045054i \\
 0.853352 - 0.047654i \\
 0.071299 + 0.058623i
\end{array}
} \right) .
\label{lbl:eqn:solutions:A}
\end{equation}

\section{Conclusions} 
\label{lbl:sec:conclusions}

In this paper we proposed quantum circuit based on qutrirs and allowing for solving SLE with three unknown variables. Presented circuit can be extended to the cases of bigger systems of linear equations but, similarly to the quantum cirucits based on qubits, where the number of unknown variables will be equal to $3^n$ (for circuits based on qubits we have $2^n$).

The presented circuit can also be  generalized to those based on qudits, which allows to define very compact quantum circuits necessary for solving SLE. 
Quantum circuits for such cases will contain only three wires and the number of unknown variables will be the same as the level of used qudits. 
Another advantage of our proposal is fact that we can use Hadamard gates instead of performing Fourier transform. However, for this case the rotation gate requires more parameters (but no more than number of unknown variables).

Unfortunately, if the eigenvalues are not known, the parameters of rotation gate necessary for finding the solution of given SLE should be chosen manually. This requires nearly continuously change of values $r_1$, $r_2$, $r_3$ in range from $-1$ to $1$ what is technically very hard to perform in physical realizations at the current level of technology. The implementation of qudits also is another difficult task to physical realisation with nowadays technology.

Another problem is influence of the noise on the both: phase estimation step and simulation of the non-unitary operation with use of rotation gate (discussion of such problem can be found for example in \cite{Gawron2012}). Nevertheless, we believe that the circuit proposed in this paper is interesting enough to be worth of future investigation and can be implemented in physical systems in the near future.


\section*{References}

\begin{thebibliography}{10}

\bibitem{Harrow2009}
Harrow A W, Hassidim A and Lloyd S 2009 {\it Phys. Rev. Lett.} \textbf{103} 150502

\bibitem{Barz2013}
Barz S, Kassal I, Ringbauer M, Lipp Y O, Dakic B, Aspuru-Guzik A and Walther P 2013  {\it Preprint} arXiv:1302.1210

\bibitem{Cai2013}
Cai X D, Weedbrook C, Su Z E, Chen M C, Gu M, Zhu M J, Li L, Liu, N L, Lu C Y and Pan J W 2013 {\it Phys. Rev. Lett.} \textbf{110}  230501

\bibitem{Cao2012}
Cao Y, Daskin A, Frankel S and Kais S 2012 {\it Molecular Phys.: An Inter. J. at the Interface Between Chem. and Phys.} {110}(2012)(15-16) p.1675--80


\bibitem{Gawron2012}
Gawron P, Klamka J and Winiarczyk R 2012 {\it Int. J. Appl. Math. Comput. Sci.}  {22}(2012){2} p.493--9

\end{thebibliography}
\end{document}